\documentstyle[12pt]{article}
\newcommand{\Frac}[2]{\frac{\displaystyle #1}{\displaystyle #2}}
\begin{document}
\pagestyle{empty}
\input{Feynman}
\begin{titlepage}
\today
\begin{center}
\hfill FTUV/93-16\\
\hfill IFIC/93-7\\
\vspace{3cm}
{\Large \bf Parameter--free calculation of $D \rightarrow P P$
\vspace*{0.3cm} \\ in a
weak--gauged  \vspace*{0.4cm} \\   $U(4)_L \otimes
U(4)_R$ Chiral Lagrangian Model} \\
\vspace{2cm}
{\large F.J. Botella, S. Noguera, J. Portol\'es} \\
\vspace{1cm}
Departament de F\'{\i}sica Te\`orica and I.F.I.C. (Centre Mixte
Universitat de Val\`encia--C.S.I.C.) \\
E-46100 Burjassot (Val\`encia, Spain) \\
\vspace{2cm}
\begin{abstract}
We analyze the  decay modes of charmed
mesons into two--pseudoscalars in a weak--gauged $U(4)_L \otimes
U(4)_R$ chiral lagrangian
model. The calculation is free of unknown
parameters and only requires as inputs the masses of pseudoscalars
and the decay constant of the pion.
The general pattern of the results at leading order compares reasonably
well with the experimental data.
\end{abstract}
\end{center}
\end{titlepage}
\newpage
\pagestyle{plain}
\pagenumbering{arabic}
\vspace{5cm}

\hspace{0.5cm}Non--leptonic weak decays have always been a challenge
for the Standard Theory.
In the last years the amount of data on non--leptonic decays
of charmed mesons have motivated to many authors the study of some theoretical
prescriptions for these processes. The point have always been the same: take
the effective four quark operators that allow the weak process and
try to implement QCD on them. The perturbative regime of strong
interactions is easy to handle in a standard way, mainly by using
the appropriate matching
conditions \cite{a}.
 But for non--perturbative effects, which are important in hadrons,
 there has been some freedom in the
election of assumptions in order to calculate hadron matrix elements.
The most important are the factorization model \cite{1} justified by the
$1/N_c$ expansion \cite{2}, the old vector--dominance model \cite{3},
the bosonization of currents in a chiral approach \cite{4}
and the more well established QCD sum rules method \cite{5}. The results
of these works are good but have support on some, more or less,
``{\em ad--hoc}" assumptions over free parameters and methods.
Non--leptonic weak decays of charmed mesons joint together the absence
of big problems in the phenomenological analysis and parameterization
with a poor knowledge about the treatment of hadronic matrix elements.
\par
In this letter we present the analysis of decays
of charmed mesons into two--pseudoscalar ($ D \rightarrow P P$)
in a weak--gauged $U(4)_L \otimes U(4)_R$ chiral
lagrangian model we have presented elsewhere \cite{6}. In a latter
publication we will present the analysis of the more involved
three body decays ($D \rightarrow P P P$) \cite{7}.

\par
The basic idea in our model is to solve the strong interaction
by using a reasonable Chiral Lagrangian and then introduce the weak
interactions through the requirement of local gauge invariance under
the electroweak group $SU(2)_L \otimes U(1)_Y$ over the hadronic
(in our case mesonic) degrees of freedom. In order to have two
doublets under $SU(2)_L$ we are forced to enlarge the Chiral Lagrangian,
so we have chosen a $U(4)_L \otimes U(4)_R$ Chiral Lagrangian. In
particular we have taken a linear realization in order to preserve
predictability as much as possible. This model gets naturally not
only all the symmetries of the two--generation Standard Theory but
also its symmetry breaking patterns.
\vspace*{1cm} \\

\hspace{0.5cm}We refer the reader to the references [7] for a
detailed explanation of our model. Here we give a brief outline.
\par
Our lagrangian is:
\begin{equation}
       {\cal L} = {\cal L}_{mesons} + {\cal L}_{Higgs}
                  + {\cal L}_{HM}  + \ldots
\end{equation}
where the dots are short for pure gauge boson terms.
Here ${\cal L}_{mesons}$ contains the strong interaction between mesons
and the couplings of mesons to the gauge bosons.
We have a set of $16$ scalars and $16$ pseudoscalars fields which
we assign to the
$(4,\bar{4}) \oplus (\bar{4},4)$ representation of the chiral
$U(4)_L \otimes U(4)_R$ group. We denote the meson matrix by
$U = \Sigma + i \Pi$, where $\Sigma$ is the scalar and $\Pi$
the pseudoscalar matrices of fields.
The explicit expression for the pseudoscalar matrix is
\begin{eqnarray*}
      \Pi & = & \left( \begin{array}{cccc}
                   \frac{1}{2}\eta_{\circ} +
                   \frac{1}{\sqrt{2}}\pi^{\circ} + & &
                   &  \\ & \pi^{+} & K^{+} & \bar{D^{\circ}} \\
                   \frac{1}{\sqrt{6}}\eta_8 + \frac{1}{\sqrt{12}}
                   \eta_{15} & & & \\
                   & & & \\ & & & \\
                    & \frac{1}{2}\eta_{\circ} -
                   \frac{1}{\sqrt{2}}\pi^{\circ} + & & \\
                   \pi^{-} & & K^{\circ} & D^{-} \\
                   & \frac{1}{\sqrt{6}}\eta_8 +
                   \frac{1}{\sqrt{12}}\eta_{15} & & \\
                   & & & \\ & & & \\
                   & & \frac{1}{2}\eta_{\circ} -
                   \frac{2}{\sqrt{6}}\eta_8 + & \\
                   K^{-} & \bar{K^{\circ}} & & D_s ^{-} \\
                   & & \frac{1}{\sqrt{12}}\eta_{15} & \\
                   & & & \\ & & & \\
                   D^{\circ} & D^{+} & D_s ^{+} &
                   \frac{1}{2}\eta_{\circ} -
                   \frac{3}{\sqrt{12}}\eta_{15}
                   \end{array}
             \right)
\end{eqnarray*}
\begin{equation}
         \;
\end{equation}

A similar matrix can be written for the scalar mesons. Our notation
for scalar mesons is $\sigma_{\circ},\sigma_8, \sigma_{15}, \sigma^+,
\sigma_3, \kappa, \delta, \delta_s$ instead of $\eta_{\circ},
\eta_8,\eta_{15}, \pi^+, \pi^{\circ}, D $ and $D_s$ respectively.
\par
With these definitions ${\cal L}_{mesons}$ is

\begin{equation}
      {\cal L}_{mesons} = \frac{1}{2} Tr[(D^{\mu}U')^{\dagger}
                                         (D_{\mu}U')] -
                           V_{chiral}(U)
\end{equation}
where $V_{chiral}$ is the chiral potential
\begin{eqnarray}
      V_{chiral}(U) & = & - \mu_{\circ}^2 \; Tr(U^{\dagger}U) +
                            \nonumber \\*
                    &   &
                            \mu_{\circ}^2 \; [  a \;
                                 Tr(U^{\dagger}U)^2  +
                                        b \; (Tr(U^{\dagger}U))^2  +
                                        c \; (det U + det U^{\dagger}
                                            )  ]
\end{eqnarray}
with $\mu_{\circ}^2 > 0$ in order to develop spontaneous breaking
of chiral symmetry.
The covariant derivative is:
\begin{equation}
      D_{\mu} U' = \partial_{\mu} U' -
                        i g \vec{T} \cdot \vec{W_{\mu}} U' -
                        i g' Y_L B_{\mu} U' +
                        i g' U' Y_R B_{\mu}
\end{equation}
with
\begin{equation}
              U' = S U S^{\dagger}
\end{equation}
Here $\vec{W_{\mu}}$ and $B_{\mu}$ are the gauge bosons related with
the $SU(2)_L$ and the $U(1)_Y$ groups. The $\vec{T}$ matrices are the
$SU(2)$ generators and $Y_L$ and $Y_R$ are the left and right
hypercharges. The matrix $S$ will be the Cabibbo rotation once
the $SU(2)_L \otimes U(1)_Y$ symmetry gets spontaneously  broken.
With our definitions for $U$ we have
\begin{equation}
\begin{array}{cc}
      \vec{T} = \frac{1}{2} \left( \begin{array}{cc}
                                   \vec{\tau} & 0 \\
                                   0 & \tau^1 \vec{\tau} \tau^1
                                   \end{array} \right)
      &
      Y_L = \frac{1}{6} I_{4\times4} \\    \\
      Y_R = \frac{1}{3} \left( \begin{array}{cccc}
                               2 &  &  &  \\
                                 & -1 &  &  \\
                                 &  & -1 &  \\
                                 &  &  & 2
                               \end{array} \right)
      &
      S = \left( \begin{array}{cccc}
                 1 & 0  & 0  & 0 \\
                 0 & \cos \theta_c & \sin \theta_c & 0 \\
                 0 & - \sin \theta_c & \cos \theta_c & 0 \\
                 0 & 0 & 0 & 1
                 \end{array} \right)
\end{array}
\end{equation}
with $\vec{\tau}$ the usual Pauli matrices and $\theta_c$ the
Cabibbo angle. The charge operator is $Q = Y_R = Y_L + T_3$.
\par
In (1) ${\cal L}_{Higgs}$ is the usual lagrangian for the minimal model
of Higgs of the Standard Theory. ${\cal L}_{HM}$, finally, is a
Higgs--mesons coupling term which will give masses to the mesons
 after the spontaneous symmetry breaking of the weak symmetry. In
order to write in a compact form this term we introduce the
usual Higgs doublet as a $4 \times 4$ matrix:
\begin{equation}
     H = \left( \begin{array}{cccc}
                \frac{1}{\sqrt{2}} (\psi - i \chi) &
                s^{+} & 0 & 0 \\
                - s^{-} & \frac{1}{\sqrt{2}} (\psi + i \chi) &
                0 & 0 \\
                0 & 0 & \frac{1}{\sqrt{2}} (\psi + i \chi) &
                - s^{-} \\
                0 & 0 & s^{+} &
                \frac{1}{\sqrt{2}} (\psi - i \chi)
                \end{array}  \right)
\end{equation}
We consider only the simplest local gauge invariant term

\begin{equation}
     {\cal L}_{HM} = Tr (A S^{\dagger} H^{\dagger} S U + h.c.)
\end{equation}
where $S$ is given in (7) and the most  general form of $A$
assuming isospin symmetry is
\begin{equation}
     A = \left( \begin{array}{cccc}
                \alpha & & & \\
                 & \alpha & & \\
                 & & \gamma & \\
                 & & & \delta
                \end{array} \right)
\end{equation}
${\cal L}_{HM}$ breaks explicitly the $SU(4) \otimes SU(4)$ chiral
symmetry. Through the spontaneous breaking of the chiral and weak
symmetry the matrices of mesons and Higgs get a non--zero vacuum
expectation value
\begin{equation}
\begin{array}{cc}
\langle \circ | U | \circ \rangle \equiv F = \Frac{1}{\sqrt{2}}
\left( \begin{array}{cccc}
        f_{\alpha} & & & \\
        & f_{\alpha} & & \\
        & & f_{\gamma} & \\
        & & & f_{\delta}
        \end{array} \right) , \; \; \;  &
\langle \circ | H | \circ \rangle \equiv  \Frac{1}{\sqrt{2}} \phi_{\circ}
I_{4 \times 4}
\end{array}
\end{equation}
Therefore from (9) we get
\begin{equation}
{\cal L}_{HM} = \Frac{\phi_{\circ}}{\sqrt{2}} Tr ( A ( U + U^{\dagger})) +
Tr ( A S^{\dagger} \tilde{H}^{\dagger} S U + h.c.)
\end{equation}
where
\begin{equation}
\tilde{H} = H - \langle \circ | H | \circ \rangle
\end{equation}
We note that the lagrangian (9)
 transforms under chiral $SU(4) \otimes SU(4)$ as the
$(4, \bar{4}) \oplus (\bar{4}, 4)$ representation and, therefore,
the first term in (12) is similar to
the usual explicit breaking of chiral symmetry.
\par
It is of interest for our study to analyze the limit in which the
masses of the scalars are very big in comparison with the ones of the
pseudoscalars. This
fact is equivalent to take the $\mu_{\circ}^2 \rightarrow \infty$
limit but keeping $\mu_{\circ}^2 c \equiv c'$ constant in order
to maintain the $\eta_{\circ}$ mass finite \cite{Witten}. In this case
$F$ in (11) is
\begin{equation}
F = \Frac{1}{\sqrt{2}} f_{\circ} I_{4 \times 4}
\end{equation}
with
\begin{equation}
f_{\circ}^2 = \Frac{1}{a + 4 b}
\end{equation}
In this limit the mass of the singlet scalar field is proportional
to $\mu_{\circ}^2$
\begin{equation}
m_{\sigma_{\circ}}^2 = 4 \mu_{\circ}^2
\end{equation}
and the same occurs with the masses of the $15-$plet
scalar mesons which become degenerated:
\begin{equation}
m_{\sigma_{15}}^2 = 4 \mu_{\circ}^2 a f_{\circ}^2
\end{equation}
\par
The chiral potential (4) mixes the neutral pseudoscalar fields
$\eta_{\circ}, \eta_8, \eta_{15}$. The $\pi^{\circ}$ decouples from the other
neutral pseudoscalar mesons due to the isospin symmetry. In the limit
$\mu_{\circ}^2 \rightarrow \infty$ the mass matrix for the $\eta$
mesons has a simple form:
\begin{equation}
    M = \left( \begin{array}{ccc}
               \frac{1}{2} (m_D^2 + m_K^2 - 8 c' f_{\circ}^2) &
               \sqrt{\frac{2}{3}} (m_{\pi}^2 - m_K^2) &
               \frac{1}{2 \sqrt{3}} ( m_K^2 + 2 m_{\pi}^2 - 3 m_D^2 ) \\
               & & \\
               \sqrt{\frac{2}{3}} (m_{\pi}^2 - m_K^2) &
               \frac{4}{3} m_K^2 - \frac{1}{3} m_{\pi}^2 &
               \frac{\sqrt{2}}{3} (m_{\pi}^2 - m_K^2) \\
               & & \\
               \frac{1}{2 \sqrt{3}} (m_K^2 + 2 m_{\pi}^2 - 3 m_D^2) &
               \frac{\sqrt{2}}{3} (m_{\pi}^2 - m_K^2) &
               \frac{1}{2} ( 3 m_D^2 + \frac{1}{3} m_K^2 - \frac{4}{3}
               m_{\pi}^2 )
               \end{array}
         \right)
\end{equation}
If we want to retain all $\eta$ meson masses finite it is necessary
to keep $c'$ finite. We note that this limit is different of the one
taken in [7]. In the processes studied in [7] the results are
independent of the behaviour of $c$  and the $\eta$ masses in the
large $\mu_{\circ}^2$ limit and therefore all the results
of [7] remain in this new limit. But for charm decays into
$\eta, \eta'$ it is necessary to have these masses under control.
\par
As we have said we will assume $SU(2)_I$ symmetry in our calculation.
 At this level the first hadronic inputs will be the masses
$m_{\pi}$, $m_K$, $m_D$ (which absorb the parameters of the $A$
matrix (10)) and the pion decay constant (we take
$f_{\circ} = f_{\pi} = 0.093 GeV$). In the large $\mu_{\circ}^2$ limit,
once those values have been chosen the only free parameter of our
lagrangian is $c'$. The value for $c'$ is fixed from $m_{\eta},
m_{\eta'}$ and $m_{\eta_c}$. These three masses obtained from the
diagonalization of $M$ in (18) have a slow dependence on $c'$. A
reasonable range of variation for $c'$ is $c' = -33, -28$. For
these values of $c'$ the masses are $m_{\eta}=0.507 GeV, 0.497 GeV$,
$m_{\eta'} = 1.023 GeV, 0.969 GeV$ and $m_{\eta_c} = 2.701 GeV,
2.691 GeV$ respectively.

Some comments are in order about this model:
\begin{itemize}
\item[a)] The starting point of our ideas
is to believe that the symmetries of the Standard Theory are essential
to describe the weak processes of hadrons, and any model for them has
to support not only its symmetries but its symmetry breaking patterns also.
\item[b)]The GIM mechanism is naturally implemented in our scheme.
Then we have not flavour changing neutral currents and, consequently,
no tree level weak decay processes involving only neutral particles.
These processes go to one loop in strong interaction in our model.
The final purpose would be to calculate every process at this one
loop in order to include consistently both
final state interactions and processes involving neutral mesons where
there is a change of flavour.
\item[c)]The model has been tested at one--loop level in radiative
rare kaon decays with satisfactory results ($K^+ \rightarrow \pi^+
l^+ l^-, K_s \rightarrow \pi^{\circ} l^+ l^-$) in [7].
\end{itemize}
What could be considered a full calculation of the decays $D \rightarrow
PP$ in the present model is certainly a formidable task. This will include
a full one loop calculation at leading order in $G_F$ in a theory
with $16 + 16$ scalar and pseudoscalar fields. Instead, as a first
approach to the problem, we will simply present a tree level calculation,
so this time we have no prediction for $D^{\circ} \rightarrow
\bar{K}^{\circ} \pi^{\circ}, \bar{K}^{\circ} \eta, \ldots$.
Neither we have final--state--strong--interacting phases. So in this
letter we are interested in the general pattern of the experimental
results.
\vspace*{1cm} \\

\hspace*{0.5cm}The diagrams which can contribute to the general
process $D \rightarrow P_1 P_2$ are showed in Fig. 1.
To make the calculations we have chosen a non--linear $R_{\xi}$
gauge \cite{8} which peculiar realization has been discussed in
[7]. In this gauge the usual direct coupling between mesons
and $W_{\mu}$ is substituted by a  mixing between mesons and $s^+$,
 the charged Higgs.
 We can make
a comparison between our diagrams and those coming from a
quark picture: we have a {\em spectator like}
diagram as given by Fig. 1.a) and an {\em annihilation like}
diagram as given by Fig. 1.b). The last diagram only can contribute to the
decays of a charged meson.

\par
The calculation of the amplitudes is greatly involved by the
parameterization of the lagrangian. As an example we take the
process $D^{\circ} \rightarrow K^- \pi^+$ which has the contribution
of the diagram in Fig. 1.a). The weak insertion in the $\pi^+$ leg
gives a factor $i \sqrt{2} G_F \cos^2 \theta_c f_{\pi} f_{\delta_s}
m_{\delta_s}^2$
where $m_{\delta_s}$ is the mass of a $\delta_s$ meson and
$f_{\delta_s} = (f_{\delta} - f_{\gamma})/2$ in terms of the parameterization
of $F$ given in (11). For the strong vertex $D^{\circ} K^- \delta_s^+$
we find from $V_{chiral} (U)$ in (4) the expression $
\sqrt{2} \mu_{\circ}^2 ( 2 a (f_{\alpha} - f_{\gamma} - f_{\delta}) +
c f_{\alpha})$ that after a few algebra and the constraints of the lagrangian
[7]
 can be written as $(m_K^2 - m_D^2) /( \sqrt{2} f_{\delta_s})$.This is,
in fact, the origin of the differences of squared masses in the expressions
of the amplitudes showed below.
\par
A preliminar analysis of our calculation pointed out that the role of
scalar fields is not so important in $D \rightarrow P P$ ( results
are not too much sensible to the scalar masses), so we have chosen a limit in
which the scalar masses are very large in comparison with
the pseudoscalar ones (limit $\mu_{\circ}^2 \rightarrow \infty$)
as explained above.

\par
The amplitudes for the processes can be written in the following
form
\begin{itemize}
\item[a)]Cabibbo allowed.
\begin{equation}
A_{cc}(D \rightarrow P_1 P_2) = G_F \cos^2 \theta_c f_{\circ}
\tilde{A}_{cc}(D \rightarrow P_1 P_2)
\end{equation}
\item[b)]Cabibbo suppressed.
\begin{equation}
A_{cs}(D \rightarrow P_1 P_2) = G_F \sin \theta_c \cos \theta_c f_{\circ}
\tilde{A}_{cs}(D \rightarrow P_1 P_2)
\end{equation}
\item[c)]Double Cabibbo suppressed.
\begin{equation}
A_{ss}(D \rightarrow P_1 P_2) = G_F \sin^2 \theta_c f_{\circ}
\tilde{A}_{ss}( D \rightarrow P_1 P_2)
\end{equation}
\end{itemize}
where the $\tilde{A} (D \rightarrow P_1 P_2)$ factor splits up into
two possible contributions, $\tilde{A} = A_1 + A_2$ , $A_1$ comes
from diagrams in Fig. 1.a) and $A_2$ from Fig. 1.b). Results for $A_1$ and
$A_2$ are given in Tables I, II and III for different $D \rightarrow
P_1 P_2$ processes.
 For the decays with $\eta$ and $\eta'$ as final states
the factors $\lambda_{ij}$ come from the unitary matrix which defines the
eigenstates of the mass matrix (18):
\begin{equation}
\left( \begin{array}{c}
        \eta \\
        \eta' \\
        \eta_c
       \end{array}
\right) =
\left( \begin{array}{ccc}
        \lambda_{11} & \lambda_{12} & \lambda_{13} \\
        \lambda_{21} & \lambda_{22} & \lambda_{23} \\
        \lambda_{31} & \lambda_{32} & \lambda_{33}
       \end{array}
\right)
\left( \begin{array}{c}
       \eta_{\circ} \\
       \eta_8 \\
       \eta_{15}
       \end{array}
\right)
\end{equation}
As can be seen from the results the contributions of annihilation
like diagrams are suppressed over the spectator like ones
by a factor $(m_K^2 - m_{\pi}^2) / (m_D^2 - m_{K,\pi}^2)$.
\par
The numerical results for the processes are given in Tables IV, V and
VI where are compared with experimental data \cite{9} when possible
\footnote{ There are not experimental results por doubly Cabibbo
suppressed decays but a preliminar upper bound [11] for the
decay $D^{\circ} \rightarrow K^+ \pi^-$ is  $\Gamma ( D^{\circ}
\rightarrow K^+ \pi^-) < 6.3 \times 10^{-16} GeV$.}.
\par
Some comments are in order:
\begin{itemize}
\item[a)]As can be seen, where we have predictions our results
are good but for a factor two or so in almost all results.
\item[b)] We have
no prediction for decays involving only neutral mesons but if we
introduce the measured final state interactions in $K \pi$ it is
quiet easy to get a reasonable result for $D^{\circ} \rightarrow
\bar{K}^{\circ} \pi^{\circ}$ for example. As we have stressed above
this would form part of a full calculation including all effects at
one loop level.
\item[c)]There is a disagreement for the $D^+
\rightarrow \bar{K}^{\circ}
\pi^+$ decay between our result and the experimental one. We think
that the problem comes from the fact that this process is dominated
by the {\bf 84} piece of the lagrangian that we know is too big at
tree level.  The {\bf 84} representation of
$SU(4)$ carries the {\bf 27}$(\Delta I = 3/2)$ of $SU(3)$ \cite{10}
which gives all the contribution to the $K^+ \rightarrow \pi^+
\pi^{\circ}$ decay and at tree level we find a factor of $4$ over
the experimental value. A similar problem occurs for $D^+ \rightarrow
\pi^+ \pi^{\circ}$ because the final state belongs ( if we assume
$SU(3)$ symmetry ) to the {\bf 27} representation of $SU(3)$.

\item[d)]The allowed decays $D_s^+ \rightarrow K^+ \bar{K}^{\circ},
\pi^+ \pi^{\circ}$ are forbidden at tree level because isospin
symmetry excludes an annihilation--like diagram and GIM forbids a
spectator like one for the first process.
\item[e)]There are two processes $D_s^+ \rightarrow K^+ \pi^{\circ}$
and $D^+ \rightarrow K^{\circ} \pi^+$ which only receive contribution
from the diagram 1.b. This is originated by the absence of flavour
changing neutral currents. Therefore these results could be much more
bigger than the predicted values.
\end{itemize}
\vspace*{1cm}
\hspace*{0.5cm}We have showed that a model which has the
symmetries of the Standard Theory and where the strong interactions
are implemented by a linear $U(4)_L \otimes U(4)_R$ chiral lagrangian
gives a reasonable pattern for the decays of charmed mesons into
two--pseudoscalars. It is worthwhile to emphasize
that our calculation is free of unknown parameters:
 the only hadronic
inputs have been some masses of mesons and the pion decay constant.
Better results could be obtained at one loop where GIM does not
operate for neutral particles, and final state interaction could be
included.

\section*{Acknowledgements}
\hspace{0.5cm}The authors are indebted to J. Bernab\'eu for stimulating
this study.
This work has been supported
by the CICYT under grant AEN--90--0040, by DGICYT under grant
PB91--0131 and also, in part, by I.V.E.I.

\newpage
\hspace*{-0.6cm}{\Large \bf Table Captions}
 \\
 \\
{\bf Table I} Factors $A_1$ and $A_2$ for the Cabibbo allowed processes.
 \\
\\
{\bf Table II} Factors $A_1$ and $A_2$ for the Cabibbo suppressed processes.
\\
\\
{\bf Table III} Factors $A_1$ and $A_2$ for the double Cabibbo suppressed
processes.
\\
\\
{\bf Table IV} Comparison of theoretical and experimental widths
 for Cabibbo allowed decays. a) Value predicted for $c' = -33$.
b) Value predicted for $c' = -28$.
\\
\\
{\bf Table V} Comparison of theoretical and experimental widths
 for Cabibbo suppressed decays. a) and b) as in Table IV.
\\
\\
{\bf Table VI} Predictions of widths for double
Cabibbo suppressed decays. a) and b) as in Table IV.
\\
\newpage
\begin{center}
{\bf Table I}
\end{center}
$$
\begin{array}{|c|c|c|}
\hline
\hline
\multicolumn{1}{|c}{Decay} &
\multicolumn{1}{|c}{A_1} &
\multicolumn{1}{|c|}{A_2} \\
\hline
\hline
& & \\
D^{\circ} \rightarrow K^- \pi^+ & m_K^2 - m_D^2 & 0 \\
& & \\
\hline
& & \\
D^+ \rightarrow \overline{K}^{\circ} \pi^+ & m_K^2 - m_D^2 & 0 \\
& & \\
\hline
& & \\
D_s^+ \rightarrow \pi^+ \eta & (m_K^2 - m_D^2) ( \lambda_{11} -
\sqrt{\Frac{2}{3}}
\lambda_{12} - \Frac{\lambda_{13}}{\sqrt{3}} ) & 0 \\
& & \\
\hline
& & \\
D_s^+ \rightarrow \pi^+ \eta' & (m_K^2 - m_D^2) ( \lambda_{21} -
\sqrt{\Frac{2}{3}}
\lambda_{22} - \Frac{\lambda_{23}}{\sqrt{3}} ) & 0 \\
& & \\
\hline
& & \\
D_s^+ \rightarrow K^+ \overline{K}^{\circ} & 0 & 0  \\
& & \\
\hline
& & \\
D_s^+ \rightarrow \pi^+ \pi^{\circ} & 0 & 0 \\
& & \\
\hline
\end{array}
$$
\newpage
\begin{center}
{\bf Table II}
\end{center}
\vspace*{-0.3cm}
$$
\begin{array}{|c|c|c|}
\hline
\hline
\multicolumn{1}{|c}{Decay} &
\multicolumn{1}{|c}{A_1} &
\multicolumn{1}{|c|}{A_2} \\
\hline
\hline
& & \\
D^{\circ} \rightarrow \pi^+ \pi^- & m_D^2 - m_{\pi}^2 & 0 \\
& & \\
\hline
& & \\
D^{\circ} \rightarrow K^+ K^- & m_K^2 - m_D^2 & 0 \\
& & \\
\hline
& & \\
D^+ \rightarrow \pi^+ \pi^{\circ} & \Frac{1}{\sqrt{2}} (m_D^2 - m_{\pi}^2)
& 0 \\
& & \\
\hline
& & \\
D^+ \rightarrow \overline{K}^{\circ} K^+ & m_K^2 - m_D^2 & 0 \\
& & \\
\hline
& & \\
D^+ \rightarrow \pi^+ \eta & (m_D^2 - m_{\pi}^2) ( \lambda_{11} + \Frac{1}{2}
\sqrt{\Frac{2}{3}} \lambda_{12} - \Frac{\lambda_{13}}{\sqrt{3}} ) & 0 \\
& & \\
\hline
& & \\
D^+ \rightarrow \pi^+ \eta' & (m_D^2 - m_{\pi}^2) ( \lambda_{21} + \Frac{1}{2}
\sqrt{\Frac{2}{3}} \lambda_{22} - \Frac{\lambda_{23}}{\sqrt{3}} ) & 0 \\
& & \\
\hline
& & \\
D_s^+ \rightarrow K^{\circ} \pi^+ & m_D^2 - m_{\pi}^2 & m_{\pi}^2 - m_K^2 \\
& & \\
\hline
& & \\
D_s^+ \rightarrow K^+ \pi^{\circ} & 0 & \Frac{1}{\sqrt{2}} ( m_{\pi}^2
- m_K^2) \\
& & \\
\hline
& & \\
D_s^+ \rightarrow K^+ \eta & (m_K^2 - m_D^2) ( \lambda_{11} -
\sqrt{\Frac{2}{3}}
\lambda_{12} - \Frac{\lambda_{13}}{\sqrt{3}} )
 & (m_{\pi}^2 - m_K^2 ) ( \lambda_{11} - \Frac{1}{2}
\sqrt{\Frac{2}{3}} \lambda_{12} +
\Frac{\lambda_{13}}{\sqrt{3}} ) \\
& & \\
\hline
& & \\
D_s^+ \rightarrow K^+ \eta' & (m_K^2 - m_D^2) ( \lambda_{21} -
\sqrt{\Frac{2}{3}}
\lambda_{22} - \Frac{\lambda_{23}}{\sqrt{3}} ) &
(m_{\pi}^2 - m_K^2 ) ( \lambda_{21} - \Frac{1}{2} \sqrt{\Frac{2}{3}}
 \lambda_{22} +
\Frac{\lambda_{23}}{\sqrt{3}} ) \\
& & \\
\hline
\end{array}
$$
\newpage
\begin{center}
{\bf Table III}
\end{center}
$$
\begin{array}{|c|c|c|}
\hline
\hline
\multicolumn{1}{|c}{Decay} &
\multicolumn{1}{|c}{A_1} &
\multicolumn{1}{|c|}{A_2} \\
\hline
\hline
& & \\
D^{\circ} \rightarrow K^+ \pi^- & m_D^2 - m_{\pi}^2 & 0 \\
& & \\
\hline
& & \\
D^+ \rightarrow K^{\circ} \pi^+ & 0 & m_K^2 - m_{\pi}^2 \\
& & \\
\hline
& & \\
D^+ \rightarrow K^+ \pi^{\circ} & \Frac{1}{\sqrt{2}} (m_{\pi}^2 - m_D^2) &
\Frac{1}{\sqrt{2}} (m_K^2 - m_{\pi}^2) \\
& & \\
\hline
& & \\
D^+ \rightarrow K^+ \eta & (m_D^2 - m_{\pi}^2) ( \lambda_{11} + \Frac{1}{2}
\sqrt{\Frac{2}{3}} \lambda_{12} - \Frac{\lambda_{13}}{\sqrt{3}} ) &
(m_K^2 - m_{\pi}^2) ( \lambda_{11} - \Frac{1}{2} \sqrt{\Frac{2}{3}}
 \lambda_{12} +
\Frac{\lambda_{13}}{\sqrt{3}} ) \\
& & \\
\hline
& & \\
D^+  \rightarrow K^+ \eta' & (m_D^2 - m_{\pi}^2) ( \lambda_{12} + \Frac{1}{2}
 \sqrt{\Frac{2}{3}} \lambda_{22} - \Frac{\lambda_{23}}{\sqrt{3}} ) &
(m_K^2 - m_{\pi}^2) ( \lambda_{21} - \Frac{1}{2} \sqrt{\Frac{2}{3}}
 \lambda_{22} +
\Frac{\lambda_{23}}{\sqrt{3}} ) \\
& & \\
\hline
& & \\
D_s^+ \rightarrow K^+ K^{\circ} & m_D^2 - m_{\pi}^2 & 0 \\
& & \\
\hline
\end{array}
$$
\newpage
\begin{center}
{\bf Table IV}
\end{center}
$$
\begin{array}{|c|c|c|}
\hline
\hline
\multicolumn{1}{|c}{Decay} &
\multicolumn{1}{|c}{\Gamma_{exp} \times 10^{14} \; \, (GeV)} &
\multicolumn{1}{|c|}{\Gamma_{theor} \times 10^{14} \; \, (GeV)} \\
\hline
\hline
& & \\
D^{\circ} \rightarrow K^- \pi^+ &  5.7 \pm 0.3 & 11 \\
& & \\
\hline
& & \\
D^+ \rightarrow \overline{K}^{\circ} \pi^+ & 1.6 \pm 0.2 & 11 \\
& & \\
\hline
& & ^{a)}  4.3 \\
D_s^+ \rightarrow \pi^+ \eta & 2.2 \pm 0.6 & \\
& & ^{b)}  3.8 \\
\hline
& & ^{a)} 4.2 \\
D_s^+ \rightarrow \pi^+ \eta' & 5.4 \pm 1.8 & \\
& & ^{b)} 4.8 \\
\hline
& & \\
D_s^+ \rightarrow K^+ \overline{K}^{\circ} & 4.1 \pm 1.0 & 0 \\
& & \\
\hline
& & \\
D_s^+ \rightarrow \pi^+ \pi^{\circ} & - & 0 \\
& & \\
\hline
\end{array}
$$
\newpage
\begin{center}
{\bf Table V}
\end{center}
$$
\begin{array}{|c|c|c|}
\hline
\hline
\multicolumn{1}{|c}{Decay} &
\multicolumn{1}{|c}{\Gamma_{exp} \times 10^{15} \; \, (GeV)} &
\multicolumn{1}{|c|}{\Gamma_{theor} \times 10^{15} \; \, (GeV)} \\
\hline
\hline
& & \\
D^{\circ} \rightarrow \pi^+ \pi^- & 2.6 \pm 0.3 & 6.9 \\
& & \\
\hline
& & \\
D^{\circ} \rightarrow K^+ K^- & 6.4 \pm 0.6 & 5.3 \\
& & \\
\hline
& & \\
D^+ \rightarrow \pi^+ \pi^{\circ} &  < 3.3  & 3.5 \\
& & \\
\hline
& & \\
D^+ \rightarrow \overline{K}^{\circ} K^+ &  4.5 \pm 1.1  & 5.2 \\
& & \\
\hline
& & ^{a)} 1.8 \\
D^+ \rightarrow \pi^+ \eta   & 4.1 \pm 1.4 & \\
& & ^{b)} 1.9 \\
\hline
& & ^{a)} 0.67 \\
D^+ \rightarrow \pi^+ \eta' &   < 4.9  & \\
& & ^{b)} 0.63 \\
\hline
& & \\
D_s^+ \rightarrow K^{\circ} \pi^+ & < 8.8 & 5.5 \\
& & \\
\hline
& & \\
D_s^+ \rightarrow K^+ \pi^{\circ} & - & 0.013 \\
& & \\
\hline
& & ^{a)} 2.1 \\
D_s^+ \rightarrow K^+ \eta & - & \\
& & ^{b)} 1.8 \\
\hline
& & ^{a)} 2.4 \\
D_s^+ \rightarrow K^+ \eta' & - & \\
& & ^{b)} 2.6 \\
\hline
\end{array}
$$
\newpage
\begin{center}
{\bf Table VI}
\end{center}
$$
\begin{array}{|c|c|}
\hline
\hline
\multicolumn{1}{|c}{Decay} &
\multicolumn{1}{|c|}{\Gamma_{theor} \times 10^{16} \; \, (GeV)} \\
\hline
\hline
& \\
D^{\circ} \rightarrow K^+ \pi^- & 3.3 \\
& \\
\hline
& \\
D^+ \rightarrow K^{\circ} \pi^+ & 0.014 \\
& \\
\hline
& \\
D^+ \rightarrow K^+ \pi^{\circ} & 1.5 \\
& \\
\hline
& ^{a)}  0.83  \\
D^+ \rightarrow K^+ \eta & \\
& ^{b)}  0.91 \\
\hline
& ^{a)} 0.43 \\
D^+ \rightarrow K^+ \eta' & \\
& ^{b)} 0.41 \\
\hline
& \\
D_s^+ \rightarrow K^+ K^{\circ} & 3.0 \\
& \\
\hline
\end{array}
$$

\newpage
\hspace{-0.6cm}{\Large \bf Figure Captions} \\
    \\
    \\
{\bf Figure 1}: Diagrams for the processes $D \rightarrow P P$:
                a)  spectator like,
                b) annihilation like. $\Sigma$ is a scalar
                 meson. \\

\newpage
\begin{figure}[h]
   \begin{Feynman}{100,60}{30,0}{1}
   \put(30,35){\fermionright}      \put(35,40){$D$}
   \put(60,35){\fermionrighthalf}  \put(65,40){$\Sigma$}
   \put(76,35){\circle{2}}
   \put(77,35){\gaugebosonrighthalf}   \put(85,40){$s^{+}$}
   \put(93,35){\circle{2}}
   \put(95,35){\fermionrighthalf}  \put(100,40){$P_1$}
   \put(60,35){\fermiondr}        \put(80,20){$P_2$}
   \end{Feynman}
\end{figure}
\vspace*{-1cm}
\begin{center}
{\bf Figure 1.a)}
\end{center}
\begin{figure}[h]
   \begin{Feynman}{100,70}{30,0}{1}
   \put(30,35){\fermionrighthalf}      \put(35,40){$D$}
   \put(46,35){\circle{2}}             \put(55,40){$s^+$}
   \put(48,35){\gaugebosonrighthalf}   \put(71,40){$\Sigma$}
   \put(64,35){\circle{2}}
   \put(65,35){\fermionrighthalf}
   \put(95,50){\fermionur}          \put(97,45){$P_1$}
   \put(80,35){\fermiondr}        \put(95,25){$P_2$}
   \end{Feynman}
\end{figure}
\vspace*{-1cm}
\begin{center}
{\bf Figure 1.b)}
\end{center}

\end{document}